\documentclass[prl,showpacs,twocolumn,superscriptaddress]{revtex4}
\usepackage{graphicx}
\usepackage{dcolumn}
\usepackage{bm}
\usepackage{amsmath}
\usepackage{amsfonts}
\begin{document}

\title{Unitarity and Entropy Change in Exclusive Quark Combination
Models}
\author{Yi Jin}
\affiliation{Department of Physics,  University of Jinan, 250022}
\author{Shi-Yuan Li}
\affiliation{School of Physics, Shandong University, 250100}
\author{Zong-Guo Si}
\affiliation{School of Physics, Shandong University, 250100}
\author{Tao Yao}
\affiliation{School of Physics, Shandong University, 250100}

\date{\today}

\begin{abstract}
 Unitarity of the
  combination model is formulated and
  demonstrated to
play the key r\^{o}le  that guarantees
the non-decrease  of  the entropy in the exclusive combination process.
\end{abstract}


\maketitle


Quark Combination Model (QCM)  was proposed in early seventies of  20th century
\cite{Anisovich:1972pq,bjorken} to describe the multi-production
process in high energy collisions, based on the constituent quark model of
hadrons. 
Various versions of     QCM  succeed
in explaining many data (for a review, see
\cite{jinyi09}, with a brief list of references therein),  and
recently  those  in central gold-gold collisions
at the Relativistic Heavy Ion Collider (RHIC) \cite{sevqcm,sd1} where
 several `unexpected' phenomena observed
 lay difficulties for other hadronization mechanisms \cite{corexp}.
Common of all the hadronization models,
QCM only responds to describe
the non-perturbative QCD phase:  The produced
`partons' in various collisions turned into constituent quarks (including both
quarks anti-quarks, same in the following);  then these
quarks combined into hadrons \footnote{Other hadronization models
have corresponding issues in different languages,  e.g.,  Lund string
 \cite{lund}.}. Especially,  the  combination process is just
the `realization' of confinement within the QCM framework.
Each kind of QCM distinguishes from others by its special
way of combination.
Without digging into details of any special kind of QCM,
one  easily figures out  two principles which it
 must respect:
 First of all, energy-momentum conservation is the
  principle law of physics,  reflecting   the basic symmetry  space-time displacement invariance.
  The models must precisely (as precisely as possible, in practice)
transfer the energy and momentum of the parton system into the constituent quark system
and then the hadron system.
Second, 
for a colour-singlet separated system,
 it is necessary to let  all the constituent quarks created in the model
combined into hadrons,
 or else there are  free quarks with
non-zero mass and energy, which obviously contradicts to any
observations that suggest confinement.  Moreover, these free quarks take
away energy and momentum, hence make danger of the
energy-momentum conservation.  This second principle is referred as
Unitarity of the relevant model.
These two principles are closely connected, with the
first one 
the natural result of the second one.



To formulate the unitarity of QCM,
 we formally  introduce  the
unitary time-evolution operator $U$ to describe the combination
process \cite{ouruni}
\begin{equation}
\label{ueq1}
\sum_h |<h|U|q>|^2=<q|U^+U|q>=1.
\end{equation}
 The quark state
 $|q>$,  describes a colour-singlet quark system,
  and the corresponding hadron state $|h>$ describes  the hadron system. The
matrix element $U_{hq}=<h|U|q>$ gives the transition amplitude.
The energy-momentum conservation is inherent, by the natural commutation
 condition  $[U,H]=[U,\bf P]=0$, with $H, \bf P$ the energy and
momentum operator of the system. This is just the
confinement which says that the total probability for the colour-singlet quark system
to transit to all kinds of hadron is exactly 1,
and agrees with the fact that all the quark states and the hadron
states are respectively two complete sets of bases of the same   Hilbert space
\footnote{This is very natural,  if one adopts that  QCD is really the uniquely
correct theory for the hadron physics, with its effective Hamiltonian $H_{QCD}$.
Then all the hadron states with definite energy-momentum should be its eigenstates and expand the  Hilbert space of states (though we do not know how to solve $H_{QCD}$ mathematically).
 While  a model  is proposed  in language of constituent
quarks which composite the hadrons, all of the quark states with definite energy-momentum should be eigenstates of the same $H_{QCD}$
(Here we consider constituent quark model, and ignore the
rare probability of exotic hadrons like glueball, hence need not consider gluon states).
So these two sets of bases are of different {\it representations},
as is more easy to be recognized if one imagines that  all
the wave functions of hadrons  are written  in terms of
 quark states in some special framework of quark models and
 notices that the planer wave function 
  as well as other
special functions (bound state wave functions) are all possible to be complete bases
 for a definite functional space, mathematically. },
i.e.,  $\sum |q><q|= \sum |h><h|=1$ for the colour-singlet system.
Recently \cite{ouruni},  it is observed that, in a combination model which respects unitarity, the production of exotic hadrons is naturally suppressed. In this paper, we would like to further sharpen the  key r\^{o}le of unitarity in exclusive QCM, by
its application to guarantee  the non-decreasing of entropy 
in the combination process for a colour-singlet separated system.

 It comes out that a  third principle need  to be checked  once   QCM finds its
irreplaceable  application  in high energy  central nuclear collisions
\cite{sevqcm,sd1,corexp},  
 where this model is employed to 
describe the hadronization (freeze-out) of
 a special system,   a bulk of quark-gluon matter (QGM) with
 non-zero density and temperature.
The entropies  of the QGM, of the constituent quark system evolved from the QGM,
as well as of the hadron system combined  from   the constituent quarks,
 can be respectively introduced.   Thus the entropy change
 should respect the second law of thermodynamics
 (under this circumstance, the energy conservation is 
 just the first law).
 This is crucial when one tries to  track back the thermal state of
 the QGM from the hadron system.
 The  combination process leads to     particle number decreased
(we will clarify that
 unitarity affirms the  invariance of the degree of freedom).
  Some na\"{i}ve
considerations often question on  the entropy  to be decreased.
However, in our opinion,   one of the key complexities
of the entropy problem is as following:
Each kind of QCM has its special set of inputs.
These inputs are not designed
  as complete as to specify the concrete value of the entropy
 which is a state function of a set of
   macroscopic quantities  needed to fix the thermal properties of the system \cite{lanl}.
   Some of these quantities have no physical relation with the combination mechanism, or the  final state
   observables predicted by QCM. 
  Therefore, they can  be introduced only for the purpose to fix entropy  without any other possible experimental tests.      Contrary to
   freely tuning these `inputs' to worship the second law, 
 we find that  
the  entropy is guaranteed to be non-decreasing by
  Unitarity and energy-momentum conservation of the combination process    {\it without any
   extra inputs}. 
   In other words, the entropy of the colour-singlet  constituent quark system is not larger than   that of the hadron system, provided that  all of the quarks are combined into the hadrons,   with the energy and momentum conservatively transferred.
  Any extra macroscopic quantity to fix the entropy  which did not appear in the model before,   will never need to appear,
  since the relation exposed here shows that the entropy change is not an
  isolated problem.

%

For feasibility, we discuss
a colour-singlet constituent quark system.  It is separated,
  belonging to micro-canonical ensemble.
  {\it All of the quarks} will be exclusively
combined into  hadrons by some `combination rules'.
This system  is the offspring  from the QGM when its interaction with
the nuclear remnant  in the beam fragmentation region is ignored.
For QCD case,  there is no free gluon radiation,
contrary to the electrodynamics case.  This simplification is easier  to be adopted.
Some  inclusive calculations and discussions based on bag model (e.g, \cite{songwang})
   treat this system  as belonging to (macro)canonical ensemble, so it is open.
    Inspection of the entropy change for such cases does not give any special indication.





  Now it is clear  that combination
process will {\it never} decrease the degree of freedom
provided that the model respects unitarity.    
 For an ideal gas system, this has guaranteed the entropy non-decreasing.
However,  only the degree of freedom is not enough to
determine the entropy for a system with  general case of interactions.
In    many kinds of
QCM,   the details of the interacting potential are not figured out,
but do not necessarily indicate the  assumption of free quark system.
The `sQGP' observation \cite{sqgp}, as well as
low energy scale QCD strong coupling  imply the contrary.
In general, the entropy $S$ is \cite{lanl}
\begin{equation}
\label{ent}
 S=-tr(\rho \ln \rho).
\end{equation}
 Here  $\rho$ is the density Matrix
   without referring to equilibrium. It can
 depend on time, as the solution of  the  Liouville Equation.
Without specifying the Hamiltonian to solve the equation,  we can formally
write by definition,
\begin{eqnarray}
\rho(t)&=&|t, i>P_i<i,t|=U(t,0) |0,i>P_i<i,0|U^\dag (t,0) \nonumber \\
   &=& U(t,0) \rho (0) U^\dag (t,0).
\end{eqnarray}
 Here $U(t,0)$ is the  time evolution operator.
Taking $\rho(0)$ as the distribution of the constituent quark system {\it just
 before combination}, while  $\rho(t)$ {\it just after}, of the hadron system,
then $U(t,0)$ is exactly the operator $U$ introduced in Eq. (\ref{ueq1}).
This is  a uniform unitary
transformation
on the Hermitian operator $\rho$, which  does not change the
 trace of $\rho \ln \rho$.
So  
the entropy holds as a constant in the combination process, same as energy and momentum.
In the following, we will give an example of QCM, whose
combination rules is  obviously unitary.

 Generally,  entropy is a  function of the energy of the system.
By requiring maximum for equilibrium, one may get the relation of entropy and energy \cite{lanl}.
 As unitarity  preserves both energy  and entropy, 
in practice, if  the relation between them
for the hadron system can be extracted  from measurement,
 it should  possibly  be extrapolated to   the constituent quark system.


As one  pursues  the above formal discussions 
to the concrete models,
i.e, to employ a concrete QCM to investigate the entropy change before and after the
combination process, unitarity is again one of the pillars.
  Two systems (states) need to be set more definite,
to `squeeze' the combination process itself:
 $A$, constituent quark system which is to be  
combined;  $B$, the hadron system, resulting just from
the combination.  This is
to eliminate other underlying  processes
before/after combination
such as  expansion of the system
or hadron interaction and decay to affect the entropy.
For generality, we need to 
 justify that the entropy can be defined for states $A$ and $B$
without the \`{a} priori assumption
 of, or waiting long time for, equilibrium of them.
Though one can expect and has many arguments to
support local equilibrium  of QGM,
it is not the state we are treating. 
The constituent quark system  bursts out
from QGM,  according to  the unsolved non perturbative dynamics.  
 Assumptions like adequate long  time
for their thermalization
and equilibrium before combination can not straightforwardly result from (possible)
equilibrium of QGM,  neither  inherent in the combination rules/methods. 
 Same case is for state $B$.



The  density matrix
$\rho(t)$ describes expectation of the probability of a system
to take a certain state from a certain Hilbert space $\bold H$
at time $t$. 
 However, given $\rho(t)$,  for each time we can
 construct a `tangent' Hilbert space $\bold H_t$
in which $\rho(t)$ is the most probable distribution and gives maximum.
Hence it effectively describes an equilibrium state
with respect to  $\bold H_t$.
  On  $\bold H_t$, the  temporal  states $A$ and $B$
 are     definite thermal  states
with definite entropies.
Unitarity  guarantees the $\bold H_t$
 for $A$ or $B$ can be taken as the same. 
Thus $A$ and $B$ can be considered as two states of a 
specific  system. 
 It is not necessary  
 to think of extra inputs to describe
 the details of them 
   in one model, which will make
 the proof of entropy non-decreasing
 in this way  a vicious circle.
$Any$ reversal process integral
\begin{equation}
\label{entch}
\Delta S = \int_A^B \frac{dQ}{T}.
\end{equation}
is just the entropy change.
So we can introduce $any$ quasi-static, reversal process to combine these two states,
which will always give the unique result.

Such ideal  processes   in fact
  embeds in many Monte-Carlo programmes   to realizing the QCM. 
Though the combination process could happen 
in a very short  time interval in reality,
in many models 
it is described and realized  step by step.  Each step only
corresponds to  the combination of a quark-antiquark pair to meson or three (anti)quarks to (anti)baryon.
 When the number of  quarks is large ($\to \infty$),
 as encountered in the  central heavy ion collisions,
each step leads to minor  (infinitesimal) change of the quark system.  
So the programme is the discrete realization of a quasi-static process.
When the programme completes  running to combine the quark system to the hadron system,
 let the programme run inversely,
we come back to the initial state without any effect residual.
So it is reversal.
Hence the integral of eq. (\ref{entch}) can be calculated by step ($i$) summation
$\sum_i \delta Q_i/T_i$  employing such programmes. 
For the programmes respect  unitarity,  no free quarks going away and no free gluon radiation, 
but all  quarks are exclusively combined into hadrons.
Ignoring QED, the whole process is {\it adiabatic}. 
 $\delta Q_i \equiv 0$,  $\forall i$, then $\Delta S=0$.
 One gets the same conclusion as above.

However,  there is one complexity for the concrete programmes in practice.
 The  fundamental  equation of thermodynamics, for each combination step
 of the programmes, is  
$\Delta U_i=\delta Q_i + \delta W_i$,
where $\delta W_i$ is the work done to the system, and $\Delta U_i$ is the energy
change.   
Since the system is  
 separated while
 ignoring interactions with nuclear remnant,  in this quasi-static process $\delta W_i \equiv 0$,  $\forall i$,
 we get
\begin{equation}
\label{sonu}
\Delta S=\sum_i \Delta U_i/T_i.
 \end{equation}
One should  check that the energy of the system 
keeps unchanged  
in each step.

At  first sight,  it seems to encounter the
  crucial `difficulty' in combination models.  Most of them
adopt the constituent quark model with fixed quark mass.
 Since hadron mass is definite,  the $2\to 1$ and $3\to 1$ processes  can
not  keep energy-momentum conservation at the same time for all
particles on-shell. Here we  employ an example QCM proposed by
Shandong Group (SDQCM) to discuss this problem.  In the calculation,
 we adopt the scenario as
fixed but tunable quark mass with only three-momentum conservation,
leave open the energy for each step. This is a simplified way to
realize the model, without inputs of the potentials among the quark
system.
The energy conservation can be considered in two aspects:

 1. In the simulation of  a definite event,
each special combination of quarks (e.g., $u \bar
d$) can correspond to several kinds of hadron states ($\pi^+,
\rho^+$...) with various masses.   $\Delta U_i$ can be either  negative or
positive.   This variety leads $\sum \Delta U_i$
($\forall i$) to be vanishing, by cancelation of $\Delta U_i$ with
different signs.  In this case the final
result of the combination can conserve energy, which is required
to give a correct description of data,
 as  mentioned in the beginning of this letter.
For a constant
temperature in the combination process, $\Delta S=0$.   
  The constant temperature is not an assumption but 
  inherent in the static, reversal and ideal process simulated by the programme.
  For all the intermediate states linking state $A$ and $B$,  there are both quarks and hadrons.  
  These states are considered as in a mixed phase,   with constant temperature.

2. In the fundamental  equation of thermodynamics, the thermal
quantities must be the ensemble-averaged one, which is calculated by
averaging among infinitely
many events  employing the Monte-Carlo programme.  Same as the  arguments  in  1.,   $\Delta U_i$ can be positive or negative for each step 
in various events.  The event-averaged
 $\Delta U_i$ is hoped
to be vanishing.
From such an observation,  the temperature behaviour  needs 
no care.

Now we demonstrate the above two points with SDQCM.
SDQCM   has been applied to various high energy collisions including
nuclear collision  \cite{jinyi09,sd1,sd2}.
In this model,
all (anti)quarks are arranged into a chain along the rapidity axis.
 The combination starts from one end to the other
 step by step.
The nearest quarks   are combined into meson or baryon.
This process is exclusive and obviously respects unitarity. It can be
modified to accommodate exotic hadrons, without breaking unitarity
\cite{ouruni}.
 This model  employs a non-saturation
potential \cite{Xie:1988wi} to  explain the constituent quark
creation and
to compensate  for  energy-momentum conservation
 during the combination process.  In programmes, the
 parameters such as constituent quark masses are introduced.
 The details of the potential are ignored.
For the most general case,   three light quark masses, and three kinds of $p_T$
distributions of these quarks, can be treated as parameters. Only the $p_T$ and
rapidity distribution  of hadrons from experiments can be taken as
constrains. 
The set of the parameter values is not unique for one set of data.
Here in calculation,  we  take the special case that up and down quarks
have the similar mass and $p_T$ distribution.

\begin{figure}
  \includegraphics[width=7.6cm, height=5.3cm]{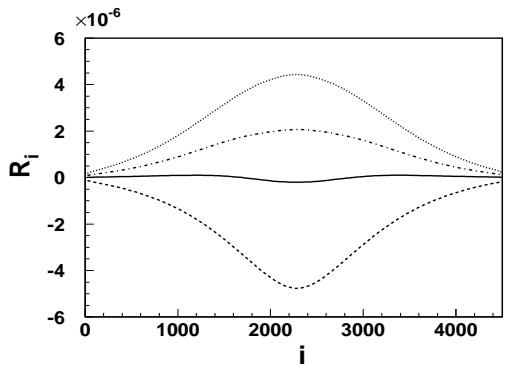}\\
  \caption{The rapidity (step) dependence for $R_{i}$.
The dashed, dotted, dot-dashed or solid line corresponds to the
total energy conservation  at the precision
 $-1\%$ ($340,500$), $1\%$ ($180, 250$), $0.5\%$ ($220,310$)
or $0.005\%$ ($260, 365$), respectively.  The numbers in the brackets are
the quark mass values in $MeV$. In each bracket,  the first one is for up and down
 quark and the second is for strange quark.}\label{fig1}
\end{figure}

To investigate the energy change of the colour-singlet  system
 in SDQCM before and
after each combination step, we define an ensemble  averaged
variable, 
\begin{equation}\label{Ri}
    R_{i}=1 - E_{i-1}/E_i.
\end{equation}
Here $E_i$ denotes energy of the colour-singlet
system  after the $ith$ combination step, $E_{i<0}=E_0$.
The ensemble average  is obtained
via event average. For the exact cancelation among events and energy conservation of a certain step, $R_i=0$.
The system in numerical calculation corresponds to the most central gold-gold collision
at center of mass frame energy (per nucleon)  of 200 $GeV$.
The  results averaged over around $10^6$ events at various precision
of total energy conservation with
 various quark mass values
 are given in Fig.\ref{fig1}.
Note that in center of mass frame of the collision,  the half of the
total step-number approximately corresponds to the mid-rapidity region.
Fig.\ref{fig1} shows the  $R_{i}$ as a function of combination step. It is clear  that one can gain high precision of the  total energy conservation,
and a vanishing $\Delta U_i$ as well. Within the corresponding precision,
Eq. (\ref{sonu}) gives $\Delta S=0$.
The effect on $R_{i}$ from modifying the invariant mass of the quark cluster
to be the corresponding hadron mass is  most
significant  in mid-rapidity  region. This comes from the fact  that
  the three-momentum for each combination step
 is exactly conserved.  Only the error (variation) of invariant mass of the
quark cluster leads to the error (variation) of the energy.
This error is   proportional to $1/\cosh y$.

In summary,  the entropy `problem' is not  isolated.
Unitarity and energy conservation
   guarantee the entropy non-decreasing
for the colour-singlet separated system of   constituent quarks combined into hadrons.
{\it Any extra inputs} are unnecessary.
Unitarity and energy conservation is
 inherent in any QCM which is applied  to get a correct
 exclusive description for
 the hadron production  data and consistent with confinement.
The relevant   discussions can be applied to other
 processes which respect unitarity and energy conservation.



~~~~~~~~~~~~~~~~~

LSY  thanks the help of
Prof. V. A. De Lorenci and Dr. W. Han  at the early stage of this work.
This part of work was done when LSY visited UNIFEI, Brasil, supported by FAPMIG.
The hospitality of UNIFEI and especially
of the De Lorenci family are greatly thanked.
Discussions with former and present members of the Shandong Group
and Dr. J. Deng are thanked.
 This work is partially supported by
NSFC with grant Nos. 10775090, 10935012, and Natural Science
Foundation of Shandong Province, China, with grant nos. ZR2009AM001, JQ200902.
YT is  supported by Independent Innovation Foundation of
Shandong University.

 \end{document}